\documentstyle[12pt,fleqn]{article}

\def\slash {\put(1,1){/}}
\def\backabit {\hskip -1.5pt}
\def\enlarge {\unitlength=1.5pt}
\def\dcircle {\displaystyle}
\def\hdots#1(#2,#3)#4{\multiput(#2,#3)(#1 3,0){#4}{\circle*{2}}}
\def\Xdots(#1,#2){
   \multiput( #1, #1)( 2.12, 2.12){#2}{\circle*{2}}
   \multiput( #1,-#1)( 2.12,-2.12){#2}{\circle*{2}}
   \multiput(-#1, #1)(-2.12, 2.12){#2}{\circle*{2}}
   \multiput(-#1,-#1)(-2.12,-2.12){#2}{\circle*{2}}
}
\def\bigcirc (#1,#2,#3){\put(#1,#2){\circle{#3}}}
\def\fcircle (#1,#2,#3){\put(#1,#2){\circle*{#3}}}
\def\hcircles(#1,#2)#3{\multiput(#1,#2)(5,0){#3}{\circle{3}}}

\def\Floop{\hdots-(-16,0)4{\thicklines\bigcirc (0,0,28)}\hdots+(16,0)4}
\def\FloopX{{\thicklines\bigcirc (0,0,28)}\Xdots(12,4)}

\def\Bloop{ 
\bigcirc (-9, 9,3)\bigcirc (-5, 12,3)\bigcirc (0, 13,3)
                  \bigcirc ( 5, 12,3)\bigcirc ( 9, 9,3)
\bigcirc (-9,-9,3)\bigcirc (-5,-12,3)\bigcirc (0,-13,3)
                  \bigcirc ( 5,-12,3)\bigcirc ( 9,-9,3) }

\def\vertex#1{\bigcirc (#1 13,0,3)\bigcirc (#1 12,5,3)
              \bigcirc (#1 12,-5,3)\hdots#1(#1 17,0)4}
\def\loop#1{{\thicklines\bigcirc (#1 13,0,12)}\hdots#1(#1 22,0)3}
\def\BloopVV{\Bloop\vertex-\vertex+}
\def\BloopVL{\Bloop\vertex-\loop+}
\def\BloopLL{\Bloop\loop-\loop+}

\def\obdots#1#2{\multiput(#1 16.5,#2 2.5)(#1 1.5,#2 2.5){6}{\circle*{2}}}
\def\obdotl#1#2{\multiput(#1 19.5,#2 5.5)(#1 1.5,#2 2.5){4}{\circle*{2}}}
\def\verteX#1{\bigcirc (#1 14,0,3)\bigcirc (#1 12,5,3)
              \bigcirc (#1 12,-5,3)\obdots#1+\obdots#1-}
\def\loopX#1{{\thicklines\bigcirc (#1 14,0,12)}\obdotl#1+\obdotl#1-}
\def\BloopVVX{\Bloop\verteX-\verteX+}
\def\BloopVLX{\Bloop\verteX-\loopX+}
\def\BloopLLX{\Bloop\loopX-\loopX+}

\def\FPloop{ \hdots-(-16,0)4\hdots+(16,0)4
\fcircle (-12, 5,1)\fcircle (-9, 9,1)\fcircle (-5, 12,1)\fcircle (0, 13,1)
\fcircle (-12,-5,1)\fcircle (-9,-9,1)\fcircle (-5,-12,1)\fcircle (0,-13,1)
\fcircle ( 12, 5,1)\fcircle ( 9, 9,1)\fcircle ( 5, 12,1)\fcircle ( 13,0,1)
\fcircle ( 12,-5,1)\fcircle ( 9,-9,1)\fcircle ( 5,-12,1)\fcircle (-13,0,1)}

\newcommand{\AmS}{{\protect\the\textfont2
  A\kern-.1667em\lower.5ex\hbox{M}\kern-.125emS}}

\def\section #1{\vskip 5mm\noindent {\bf #1}\vskip 3mm}
\def\thebibliography#1{\makeatletter\def\z@{0pt}
   \section{REFERENCES}\list{\arabic{enumi}.}
  {\settowidth\labelwidth{#1.}\leftmargin=1.67em
   \labelsep\leftmargin \advance\labelsep-\labelwidth
   \itemsep\z@ \parsep\z@
   \usecounter{enumi}}\def\makelabel##1{\rlap{##1}\hss}%
   \def\newblock{\hskip 0.11em plus 0.33em minus -0.07em}
   \sloppy \clubpenalty=4000 \widowpenalty=4000 \sfcode`\.=1000\relax}

\begin{document}

\hfill {\bf SMC-PHYS-156}

\hfill {\bf nucl-th/9709001}

\vskip 5mm
%\vskip 10mm

\noindent 
{\large  Nambu-Jona-Lasinio Model at the Next-to-Leading Order in $1/N$}
\footnote{
Invited talk presented at 
International Conference on Quark Lepton Nuclear Physics,
Osaka, Japan, May 1997.
To be published in Proceedings.
}

\vskip 3mm
\noindent 
K.~Akama$\rm^a$

\vskip 3mm
\noindent 
$\rm^a$Department of Physics, Saitama Medical College,
	Kawakado, Moroyama, Saitama \\350-04, Japan

\vskip 3mm

We derived and solved the compositeness condition 
	in the Nambu-Jona-Lasinio model
	at the next-to-leading order in $1/N$,
	and obtained the expressions 
	for the effective coupling constants
	in terms of the compositeness scale. 
In the NJL model with a scalar composite,
	the next to leading contributions are too large for $N=3$.
In the induced gauge theory with abelian gauge symmetry,
	the correction term is reasonably suppressed, while, 
	in the $SU(N_c)$ gauge theory with $N_f$ flavors of fermions,
	the correction is suppressed only for $N_f>11N_c/2$,
	complementarily to asymptotic freedom.

\section {1. INTRODUCTION}

This talk is based on the works done in collaboration with
	Takashi Hattori from Kanagawa Dental College \cite{AH}.
The Nambu-Jona-Lasinio model is a model of fermions 
	interacting through a chirally symmetric four-fermion interaction,
	through which composite bosons are formed \cite{NJL}.
This model offers us a simple field theoretical scheme
	which realizes tractable compositeness and
	spontaneously broken symmetry \cite{AT}.
It has, however, theoretical drawbacks due to its non-renormalizability.
Then the following questions are frequently asked: 
``How can we regulate 
	the badly divergent higher order contributions?"
``Can we believe the lowest order results 
	in spite of the badly divergent contributions?" 
``How should we take into account the quantum effects of 
	the composite bosons?"
The purpose of this talk is to answer these questions
	from the view points of the $1/N$ expansion \cite{1/N},
	where $N$ is the number of the fermion species.
We first note that the Nambu-Jona-Lasinio model is a special case of  
	the renormalized Yukawa model
	under the compositeness condition $Z_3=0$ \cite{cc},
	where $Z_3$ is the wave-function renormalization constant
	of the to-be-composite boson.
Then, we can calculate everything in the Nambu-Jona-Lasinio model
	by calculating the corresponding quantity 
	in the well-understood renormalized Yukawa model, 
	and after that by imposing the compositeness condition.
Here we work out the compositeness condition 
	at the next-to-leading order in $1/N$,
	and solve it to obtain the coupling constant
	in terms of the compositeness scale.

\section {2. NAMBU-JONA-LASINIO MODEL}

We consider the NJL model for the fermion 
	$\psi =\{\psi _1,\psi _2,\cdots ,\psi _N\}$ with $N$ colors
	interacting through the four fermion interaction 
	with $U(1)\times U(1)$ chiral symmetry:
\begin{eqnarray} 
  {\cal L}_{\rm NJL}=\overline \psi i\slash \partial \psi  
                    + f|\overline \psi _{\rm L}\psi _{\rm R}|^2.     
\end{eqnarray} 
In 3+1 dimensions, it is not renormalizable,
	and we assume a very large but finite momentum cutoff 
	which is taken as the compositeness scale.
This Lagrangian ${\cal L}_{\rm NJL}$ is known to be equivalent 
 	to the linearized Lagrangian \cite{aux}
\begin{eqnarray} 
  {\cal L}'_{\rm NJL}
          =\overline \psi i\slash \partial \psi 
          +(\overline \psi _{\rm L} \phi  \psi _{\rm R}+{\rm h.c.})
          -{1\over f}|\phi |^2
\end{eqnarray} 
	which is written in terms of the auxiliary field $\phi $.
Now compare this with the renormalized Yukawa model
\begin{eqnarray} 
  {\cal L}_{\rm Yukawa}
     =Z_\psi \overline \psi _{\rm r}i\slash \partial \psi _{\rm r}
     +Z_g g_{\rm r}(\overline \psi _{\rm rL} \phi _{\rm r} \psi _{\rm rR}
                         +{\rm h.c.})
     +Z_\phi |\partial _\mu \phi _{\rm r}|^2  
     -Z_\mu   \mu _{\rm r}^2|\phi _{\rm r}|^2 
     -Z_\lambda \lambda _{\rm r}|\phi _{\rm r}|^4
\end{eqnarray}
	where the quantities with the index r are the renormalized ones,
	$g$ and $\lambda $ are the effective coupling constants,
	$\mu $ is the boson mass,
	and $Z$'s are the renormalization constants.
We can see that they share the fermion kinetic term, 
	fermion-boson interaction term and the boson mass term,
	but the boson kinetic term and the four boson interaction term
	are absent in the NJL Lagrangian.
Then, if
\begin{eqnarray} 
	Z_\phi =Z_\lambda =0,\label{cc}
\end{eqnarray} 
	the Yukawa model Lagrangians reduces to the NJL model Lagrangian,
	as far as we identify
\begin{eqnarray} 
  \psi =\sqrt {Z_\psi }\psi _{\rm r},\ \ \ 
  \phi ={Z_g \over Z_\psi }g_{\rm r}\phi _{\rm r}, \ \ \ 
  f={Z_g^2g^2_{\rm r}\over Z_\psi ^2Z_\mu  \mu _{\rm r}^2}
\end{eqnarray} 
	in the Yukawa model.
The condition (\ref{cc})
	is called the compositeness condition \cite{cc}.
Thus NJL model is the special case 
	of the well-understood renormalized Yukawa model 
	specified by the compositeness condition.
The compositeness condition gives rise to relations
	among coupling constants $g_{\rm r}$, $\lambda _{\rm r}$, 
	and the cut off $\Lambda $.
If the chiral symmetry is spontaneously broken,
	they imply relations among the fermion mass $m_f$, 
	the Higgs-scalar mass $M_H$, and the cutoff $\Lambda $.
Thus we can analyze everything in the NJL model
	by investigating the well-understood Yukawa model,
	and by imposing the compositeness condition 
	on the coupling constants and masses. 
Then what is urgent is 
	to work out the compositeness condition,
	and solve it for the coupling constants.

\begin{center} 
\begin{picture}(10,10) 
\thicklines\enlarge
\put(-135,-10){
  \put( 50,3){  \Floop }
  \put( 85,0){  $\sim g_{\rm r}^2NIp^2$, }
  \put(175,3){  \hdots+(-18,0){13} 
                  \put(-5,-5){\line(1, 1){10}}
                  \put(-5, 5){\line(1,-1){10}}  }
  \put(200,0){  $=(Z_\phi -1)p^2$,}
}
\put(-135,-50){
  \put( 50,3){  \FloopX }
  \put( 85,0){  $\sim g_{\rm r}^4NI$,}
  \put(175,3){  \Xdots(0,7)
                  \put(0,-7){\line(0,1){14}}
                  \put(-7,0){\line(1,0){14}}  }
  \put(200,0){  $=(Z_\lambda -1)\lambda _{\rm r}$,}
}					
\end{picture} 
\end{center} 
\vspace{35mm}

\leftskip  10mm \rightskip  10mm \noindent 
{\bf Fig. 1}\ \ 
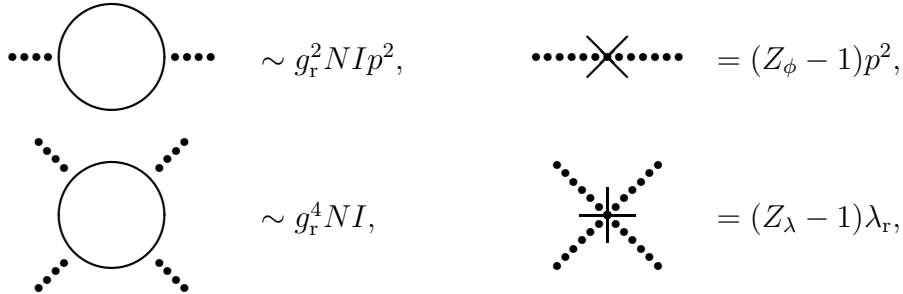
The boson self-energy part and the four-boson vertex part 
	in the lowest order in $1/N$.
	The lines 
	{\enlarge\hskip 0pt\put(5,3){\thicklines\line(1,0){21}}}\hskip 46pt 
 	and {\enlarge\hdots+(5,3)8}\hskip 46pt 
        stand for the fermion and boson propagator, respectively.

\leftskip  0mm \rightskip  0mm \vskip 5mm

For an illustration, we begin with 
	the lowest-order contributions in $1/N$ expansion.
In the Yukawa model, 
	the boson self-energy part and the four-boson vertex part 
	are given by the diagrams in Fig.\ 1 and their counter terms,
	where $I$ is the divergent integral: 
\begin{eqnarray} 
I=\cases{
     {\displaystyle {1\over 16\pi ^2}{1\over \epsilon }}
	             \ \  {\rm (dimensional\ regularization)}
		\ \left( \displaystyle \epsilon ={4-d\over 2}\right) \cr 
      \displaystyle {1\over 16\pi ^2}\log\Lambda ^2
                     \ \ \  {\rm (Pauli\ Villars\ regularization)} }
\end{eqnarray} 	
The renormalization constants $Z_\phi $ and $Z_\lambda $ should be chosen as 
\begin{eqnarray} 
  Z_\phi =1-g_{\rm r}^2NI,\ \ \ \ 
  Z_\lambda \lambda _{\rm r}=\lambda _{\rm r} - g_{\rm r}^4NI
\end{eqnarray} 
	so as to cancel out all the divergences in Fig.\ 1.
Then the compositeness condition is obtained by simply putting 
	$Z_\phi $ and $Z_\lambda $ vanishing,
\begin{eqnarray} 
    0=1-g_{\rm r}^2NI,\ \ \ \ 0=\lambda _{\rm r} - g_{\rm r}^4NI,
\end{eqnarray} 
	and it is easily solved 
	to give the expressions for the coupling constants \cite{ES}.
\begin{eqnarray} 
    g_{\rm r}^2={1\over NI},\ \ \ \ \
    \lambda _{\rm r}=\displaystyle {1\over NI}.
\end{eqnarray} 
If $\mu _{\rm r}<0$, the chiral symmetry is spontaneously broken, and
	the masses of the physical fermion and physical Higgs scalar
	are given in terms of cutoff: 
\begin{eqnarray} 
    m_f=g_{\rm r}\langle \phi \rangle 
	=\langle \phi \rangle /\sqrt {NI},\ \ \ \
    M_H=2\sqrt \lambda _{\rm r}\langle \phi \rangle 
	=2\langle \phi \rangle /\sqrt {NI}
\end{eqnarray} 
The Higgs mass is twice the fermion mass.
\begin{eqnarray} 
    2m_f=M_H
\end{eqnarray} 
These reproduce the well known results of the lowest order 
	Nambu-Jona-Lasinio model \cite{NJL}.

\vskip 10mm
\begin{center} 
\begin{picture}(10,10) 
\enlarge
\put(-120,0){ \Floop\bigcirc (-9,6,3)\bigcirc (-5,3,3)\bigcirc (0,2,3)
                    \bigcirc (5,3,3)\bigcirc (9,6,3) }
\put(-60,0){  \FloopX\bigcirc (-6,9,3) \bigcirc (-5,4,3) \bigcirc (0,2,3) 
                      \bigcirc (5,4,3) \bigcirc (6,9,3) }
\put( -0,0){  \BloopVVX} 
\put(  60,0){  \BloopVLX}
\put( 120,0){  \BloopLLX}	\label{vt2}
\end{picture} 
\end{center} 
\vspace{10mm}

\leftskip 10mm \rightskip 10mm \noindent 
{\bf Fig. 2}\ \ 
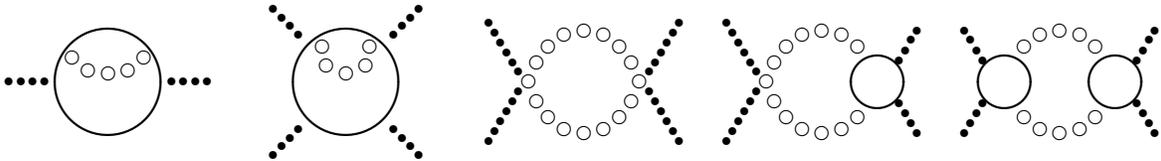
The boson self-energy part and the four-boson vertex part 
	in the next-to-leading order in $1/N$.
	The chains of tiny circles 
	stand for the infinite sum of the fermion loop insertions
	into the boson propagator.

\leftskip  0mm \rightskip  0mm \vskip 5mm
Now we turn to the next-to-leading order in $1/N$ expansion.
In the Yukawa model, 
	the boson self-energy part and the four boson vertex part 
	are given by the diagrams in Fig. 2
	and the counter terms for all the subdiagram divergences.
The renormalization constants $Z_\phi $ and $Z_\lambda $
	are calculated to be like 
\begin{eqnarray} 
	Z_\phi =1-g_{\rm r}^2NI-g_{\rm r}^2I-{1\over N}(1-g_{\rm r}^2NI)
	\log(1-g_{\rm r}^2NI)
\end{eqnarray} 
\begin{eqnarray} 
Z_\lambda \lambda _{\rm r}&=&\lambda _{\rm r} - g_{\rm r}^4NI+8g_{\rm r}^4I
       +{20(\lambda _{\rm r} - g_{\rm r}^2)^2I\over 1-g_{\rm r}^2NI}
\cr &&
 -{1\over N}\left[ 2g_{\rm r}^2(1-g_{\rm r}^2NI)+20(\lambda _{\rm r} 
                  - g_{\rm r}^2)\right] 
\log(1-g_{\rm r}^2NI)
\end{eqnarray} 
	so as to cancel out all the divergences in Fig.\ 2. 
The logarithm arises from the infinite sum 
	over the fermion loop insertions into the internal boson lines.
The compositeness condition is given 
	by putting these expressions vanishing.
Though it looks somewhat complicated at first sight,
	it can be solved by iteration 
	to give this very simple solution.
\begin{eqnarray} 
g_{\rm r}^2
	=\displaystyle {1\over NI}\left[ 1-{1\over N}
                  +O({1\over N^2})\right] ,
\ \ \ \ 
\lambda _{\rm r} 
	=\displaystyle {1\over NI}\left[ 1-{10\over N}+
               O({1\over N^2})\right] .
\end{eqnarray} 
The next-to-leading correction to the ratio of $M_H$ and $m_f$,
	which was 2 in the lowest order,
	is calculated to be:
\begin{eqnarray} 
	{M_H\over m_f}
	={g_{\rm r}\over \sqrt {\lambda _{\rm r}}}
	=2\left[ 1-{9\over 2N}+O({1\over N^2})\right] 
\end{eqnarray} 
For the case of $N=3$ of the practical interest,
	the corrections turn out to be too large,
	and the coupling constant $\lambda $ is negative,
	which implies that the Higgs potential is unstable.

For the case of $N=3$ of the practical interest,
	the corrections turn out to be too large,
	and the coupling constant $\lambda $ is negative,
	which implies that the Higgs potential is unstable.
This seems to be a serious difficulty of this model.
A way out of this difficulty is as follows.
In practical applications,
	$\psi $ is the quark and $\phi $ is the meson.
Then it is rather natural to take  
	the cutoff $\Lambda _\psi $ for the quark propagator 
	is much larger than the cutoff $\Lambda _\phi $
	for the meson propagator. 
An actual calculation shows that
	the NLO correction terms are suppressed by a factor 
\begin{eqnarray} 
	r = \log\Lambda _\phi ^2/ \log\Lambda _\psi ^2,
\end{eqnarray} 
	and the higher order terms in $1/N$ are further suppressed.

\section {3. INDUCED GAUGE THEORY}

We can apply \cite{AH} this method to the induced gauge theory \cite{ind},
	namely, the gauge theory with a composite gauge field.
It is given by the strong coupling limit 
	$f\rightarrow \infty $ \cite{Birula} of the
	vector-type four Fermi interaction model 
	for the fermion $\psi $ with the mass $m$:
\begin{eqnarray} 
   {\cal L}_{\rm 4F}
   =\overline \psi _j \left( i\slash \partial -m\right) \psi  
    -f\left( \overline \psi \gamma _\mu \psi \right) ^2,
\end{eqnarray} 
where $f$ is the coupling constant.
The Lagrangian ${\cal L}_{\rm 4F}$ is equivalent to 
\begin{eqnarray} 
 {\cal L}'_{\rm 4F}
 =\overline \psi \left( i\slash \partial -m- \slash \backabit A\right) \psi 
\end{eqnarray} 
written in terms of the vectorial auxiliary field $A_\mu $.
Then we can see that 
	this is the special case of the renormalized gauge theory
\begin{eqnarray} 
   {\cal L}_{\rm G}
   =\overline \psi _{\rm r}\left( i Z_2 \slash \partial -Z_m m_{\rm r} 
         -Z_1 e_{\rm r}\slash \backabit A_{\rm r}\right) \psi _{\rm r}
         -{ 1 \over 4 } Z_3 \left( \partial _\mu  A_{{\rm r}\nu }
                                  - \partial _\nu  A_{{\rm r}\mu } 
                            \right)  ^2 ,
\end{eqnarray} 
	specified by the compositeness condition
\begin{eqnarray} 
	Z_3=0,
\end{eqnarray} 
	where the quantities indicated by suffices r are renormalized ones,
	$e$ is the effective coupling constant, and
	$Z$'s are the renormalization constants.

\vspace{5mm}
\begin{center} 
\begin{picture}(10,10) \enlarge
  \put(-60,0){ \Floop }
  \put(  0,0){ \Floop 
               \bigcirc (-9,  6,3)\bigcirc (-5, 3,3)\bigcirc (0,2,3)
               \bigcirc ( 5,  3,3)\bigcirc ( 9, 6,3)}
  \put( 60,0){ \Floop 
               \bigcirc ( 0,-10,3)\bigcirc ( 0,-5,3)\bigcirc (0,0,3)
               \bigcirc ( 0,  5,3)\bigcirc ( 0,10,3)}
\end{picture} 
\end{center} 
\vspace{5mm}

\leftskip 10mm \rightskip 10mm \noindent 
{\bf Fig. 3}\ \ 
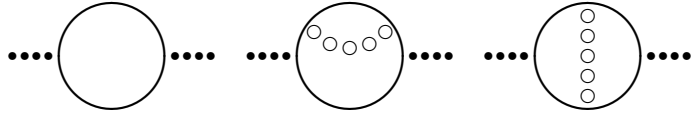
The gauge boson self-energy part in the lowest 
	and next-to-leading order in $1/N$.

\vspace{5mm}
\leftskip 0mm \rightskip 0mm 
The self-energy part of the gauge boson 
	at the leading and the next-to-leading order.
	is given by the diagrams in Fig.\ 3.
The renormalization constant $Z_3$ is chosen 
	so as to cancel out the divergences in these diagrams.
After a lengthy calculation we obtain the following expression for $Z_3$:
\begin{eqnarray} 
Z_3=1-{e_{\rm r}^2 N \over 12\pi ^2 \epsilon  } 
    -{3e_{\rm r}^2 \over 16\pi ^2 }
	\left[  1+\left(  1-{12\pi ^2\epsilon \over e_{\rm r}^2 N }\right)  
	\ln\left(  1-{e_{\rm r}^2 N \over 12\pi ^2\epsilon }\right) \right] ,
\end{eqnarray} 
where $\epsilon =(4-d)/2$ with the dimension $d$.
Then the compositeness condition $Z_3=0$ 
	is solved to give the simple solution:
\begin{eqnarray} 
	e_{\rm r}^2 = { 12 \pi ^2 \epsilon \over N }
                      \left[  1-{9\epsilon \over 4 N }\right] .
\end{eqnarray} 
The correction term $9\epsilon /4N$ is 
	naturally suppressed by the small factor $\epsilon $.
It justifies the lowest order approximation of this model
	unlike in the case of the 
	aforementioned NJL model of the scalar composite.
The origin of the suppression factor is traced back to 
	the gauge cancellation of the leading divergence 
	in the next-to-leading (in $1/N$) diagrams in Fig.\ 3.

So far we assumed that all the fermions has the same charges for simplicity.
If the charges $Q_i$ are different, the expression is modified as
\begin{eqnarray} 
	e_{\rm r}^2 ={ 12 \pi ^2 \epsilon \over \sum _j Q_j^2 }
	\left[  1-{9\epsilon \sum _j Q_j^4\over 4(\sum _j Q_j^2)^2}\right] .
\end{eqnarray} 
If we apply this to the quantum electrodynamics
	with 3 generations of quarks and leptons,
	$\epsilon $ is estimated to be 6$\times 10^{-3}$,
	which implies the next-to-leading order correction amounts 
	only to 0.1\% of the lowest order term.

Now we apply it \cite{AH} to the non-abelian induced gauge theory.
We consider the system of $N_c$ gauged colors and $N_f$ ungauged flavors. 
The starting four Fermi interaction 
\begin{eqnarray} 
   {\cal L}_{\rm 4F}=\overline \psi \left( i\slash \partial -m\right) \psi  
    -f\left( \overline \psi \lambda ^a\gamma _\mu \psi \right) ^2,
\end{eqnarray} 
	involves the internal symmetry
	expressed by the Gell-Mann matrix $\lambda ^a$($a=1,\cdots ,N_c^2-1$),
	and the auxiliary field $A_\mu ^a$ 
	in the equivalent Lagrangian (in the limit $f\rightarrow \infty $)
\begin{eqnarray} 
    {\cal L}'_{\rm 4F}
    =\overline \psi \left( i\slash \partial -m
                           -\lambda _a\slash \backabit A^a
                    \right) \psi 
\end{eqnarray} 
	also has an internal symmetry.
Then this is the special case of the renormalized gauge theory
\begin{eqnarray} 
{\cal L}_{\rm G}&=&\overline \psi _{\rm r}
\left( i Z_2 \slash \partial -Z_m m_{\rm r}-Z_1 
g_{\rm r}\lambda ^a\slash \backabit A_{\rm r}^a\right) \psi _{\rm r}
\cr &&
  -{ 1 \over 4 } Z_3 
  \left( \partial _\mu  A_{{\rm r}\nu }^a 
       - \partial _\nu  A_{{\rm r}\mu }^a 
       + Z_3^{1/2} Z_g g_{\rm r}f^{abc} A_{{\rm r}\mu }^b A_{{\rm r}\nu }^c
  \right)  ^2 ,
\end{eqnarray} 
	specified by the compositeness condition 
\begin{eqnarray} 
	Z_3=0,
\end{eqnarray} 
	where the quantities indicated by suffices r are renormalized ones,
	$g$ is the effective coupling constant, and
	$Z$'s are the renormalization constants.

\vspace{5mm}

\begin{center} 
\begin{picture}(10,10) \enlarge
  \put(-70,0){ \Floop }
  \put(  0,0){ \Floop 
               \bigcirc (-9,6,3)\bigcirc (-5,3,3)\bigcirc (0,2,3)
               \bigcirc (5,3,3)\bigcirc (9,6,3)}
  \put( 70,0){ \Floop 
               \bigcirc (0,-10,3)\bigcirc (0,-5,3)\bigcirc (0,0,3)
               \bigcirc (0,5,3)\bigcirc (0,10,3)}

  \put(-100,-35){ \BloopVV}
  \put( -35,-35){ \BloopVL}
  \put(  35,-35){ \BloopLL} 
  \put( 100,-35){ \FPloop }		
\end{picture} 
\end{center} 
\vspace{25mm}

\leftskip 10mm \rightskip 10mm \noindent 
{\bf Fig. 4}\ \ 
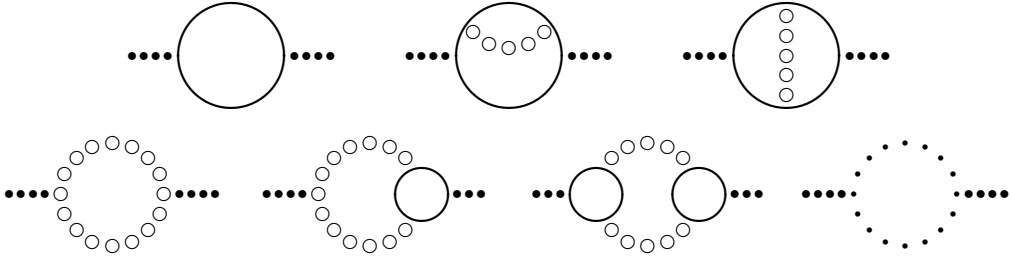
The gauge boson self-energy part in the lowest order
	and next-to-leading order in $1/N$.
	The line {\enlarge
	\fcircle (5,3,1)\fcircle (10,3,1)\fcircle (15,3,1)
        \fcircle (20,3,1)\fcircle (25,3,1)}\hskip 43pt
	stands for the Fadeev Popov ghost propagator.

\vspace{5mm}
\leftskip 0mm \rightskip 0mm 

The self-energy part of the gauge boson 
	is given by the diagrams in Fig.\ 4
	at the leading order and the next-to-leading order in $1/N_f$.
The renormalization constant $Z_3$ is chosen 
	so as to cancel out the divergences in these diagrams.
After a lengthy calculation we obtain the following expression.
\begin{eqnarray} 
Z_3 &=& 1 - { 2\over 3} N_f g_{\rm r}^2 I
         + {11\over 3} N_c g_{\rm r}^2 I
         - {\alpha \over 2} N_c g_{\rm r}^2 I 
           (1 - { 2\over 3} N_f g_{\rm r}^2 I)
\cr &&
   + {3\over 2} N_c ( {3\over 2N_f} - g_{\rm r}^2 I)
                    \ln(1 - { 2\over 3} N_f g_{\rm r}^2 I)
    +O({1\over N_f^3}).
\end{eqnarray} 
Then the compositeness condition $Z_3=0$ 
	is solved to give the simple solution:
\begin{eqnarray} 
  g_{\rm r}^2 
  = \dcircle {3\over 2N_fI}
    \left[ 1+{11N_c\over 2N_f}+O({1\over N_f^2})\right] .
\end{eqnarray} 
Though the $Z_3$ itself does depend on gauge parameter $\alpha $,
	the solution $g_{\rm r}$ to the compositeness condition $Z_3=0$ 
	does not.
This should be so 
	because the coupling constants and the compositeness scale
	are observable objects.
The solution is true only when 
	the next-to-leading contribution
	is smaller than the leading order one. 
This implies the relation 
\begin{eqnarray} 
	N_f>{11N_c\over 2}.
\end{eqnarray} 
Note that this region for $N_f$ and $N_c$ is 
	complementary to that for asymptotic freedom.
When the gauge theory is asymptotically free,
	the next-to-leading contributions are too large,
	so that the gauge bosons cannot be a composite of the above type.
And when it is asymptotically non-free, 
	the next-to-leading order contributions are reasonably suppressed,
	and the gauge boson can be interpreted as a composite.

\section {4. SUMMARY}

In summary, 
The composite dynamics of the NJL model
	is clarified by analyzing the well-understood
	renormalized Yukawa model with the compositeness condition, 
We derived and solved the compositeness condition in various models
	at the next-to-leading order,
	and obtained the expressions 
	for the effective coupling constants
	in terms of the compositeness scale. 
In the NJL model with a scalar composite,
	the corrections are $-1/N$ and $-10/N$,
	which are too large for $N=3$ of our practical interest,
	and furthermore $\lambda $ is negative,
	which implies an unstable Higgs potential.
A way out of this difficulty is to use the fact that
	the cutoff for the composite boson is much smaller than 
	the cutoff for the elementary fermion. 
On the other hand,
	in the induced gauge theory with abelian gauge symmetry,
	the correction term is $-9\epsilon /4N$, 
	which is reasonably suppressed.
In the non-abelian gauge theory, the correction is $11N_c/2N_f$.
This is true only when the next-to-leading contribution
	is smaller than the leading order one, 
	which implies that $N_f>11N_c/2$.
This condition for $N_f$ and $N_c$ is 
	complementary to that of asymptotic freedom. 
We expect that the methods and results presented here 
	will be useful in pursuing composite dynamics of nuclei and hadrons,
	as well as possible compositeness of gauge bosons.

\end{document}